\documentclass[prl,floatfix,letterpaper,twocolumn,superscriptaddress]{revtex4-1}

\usepackage[pdftex]{graphicx}
\usepackage{amssymb,amsfonts,amsmath}
\usepackage{graphics,epsfig,bbm}
\usepackage{amscd,latexsym,color}
\usepackage{amsthm}
\usepackage{accents}
\usepackage{algorithmic}
\usepackage{color}

\theoremstyle{plain} \newtheorem{thm}{Theorem}

\usepackage{times}

\newcommand{\bra}[1]{\langle{#1}|}
\newcommand{\ket}[1]{|{#1}\rangle}

\newcommand{\ketbra}[2]{|{#1}\rangle\!\langle{#2}|}

\newcommand{\op}[1]{{\hat{#1}}}
\newcommand{\sop}[1]{{\mathcal{#1}}}

\DeclareMathAlphabet{\matheu}{U}{eus}{m}{n}

\DeclareMathOperator{\tr}{tr}

\DeclareMathOperator{\id}{id}

\def\cO{\mathcal{O}}

\begin{document}

\title{Practical characterization of 
quantum devices without tomography}

\date{\today} 

\author{Marcus P. da Silva} \affiliation{ Disruptive Information
  Processing Technologies Group, Raytheon BBN Technologies, Cambridge,
  Massachusetts, 02138, USA} \affiliation{D\'epartement de Physique,
  Universit\'e de Sherbrooke, Sherbrooke, Qu\'ebec, J1K 2R1, Canada}
\author{Olivier Landon-Cardinal} \affiliation{D\'epartement de
  Physique, Universit\'e de Sherbrooke, Sherbrooke, Qu\'ebec, J1K 2R1,
  Canada} \author{David Poulin} \affiliation{D\'epartement de
  Physique, Universit\'e de Sherbrooke, Sherbrooke, Qu\'ebec, J1K 2R1,
  Canada}

\begin{abstract}
  Quantum tomography is the main method used to assess the quality of
  quantum information processing devices. However, the number of
  experimental settings and the data processing time required to
  extract complete information about a device via tomography grows
  exponentially with the device size. Part of the problem is that tomography
  generates much more information than is usually sought. Taking a
  more targeted approach, we develop schemes that enable (i)
  estimating the fidelity of an experiment to a theoretical ideal
  description, (ii) learning which description within a reduced subset
  best matches the experimental data. Both these approaches yield a
  significant reduction in resources compared to tomography. In
  particular, we demonstrate that fidelity can be estimated from a
  number of simple experiments that is independent of the system size,
  removing an important roadblock for the experimental study of larger
  quantum information processing units.
\end{abstract}

\maketitle

The building blocks for quantum computers have been demonstrated in a
number of different physical
systems~\cite{YPA+03,CLS+04,HHR+05,WRR+05,LKS+05,MSB+11}. In order to
quantify how closely these demonstrations come to the ideal
operations, the experiments are fully characterized via either {\em
  quantum state tomography}~\cite{VR89} or {\em quantum process
  tomography}~\cite{PCZ97}. An important advantage of these methods is
that they require only simple local measurements.  The main drawbacks
however are that tomography fundamentally requires both experimental
and data post-processing resources that increase exponentially with
the number of particles $n$~\cite{MRL08}.

It is important to realize that the exponential cost of tomography is
not a problem restricted to a large number of qubits. For example,
recent ion trap experiments characterizing an 8 qubit state required
10 hours of measurements, despite collecting only 100 samples per
observable~\cite{HHR+05}. Surprisingly, the post-processing of the
data obtained from these experiments took approximately a
week~\cite{Blu10}. Under similar time scales, the characterization of
a 16 qubit state would take years of measurements, and over a century
of data post-processing. This is clearly a major obstacle in the
demonstration of working quantum computers, even at sizes moderately
larger than what has been demonstrated to date.

Moreover, one of the key assumptions for the fault-tolerance theorems
of quantum computation is that the noise on elementary components does
not scale badly with the system size~\cite{AGP06}. Therefore, despite
the fact that universal quantum computation can be realized with one-
and two-qubit elementary operations, it is not sufficient to
characterize small gates---larger systems may have significant noise
contributions from correlated sources as seen in recent
experiments~\cite{MSB+11}. The characterization of multi-qubit states
and operations provides crucial
information for the verification of these assumptions, and therefore
the development of large quantum information processors.

Part of the problem with the usual approach is that tomography often
provides more information than what is truly sought. Given an
experiment that prepares a quantum state represented by a density
operator $\op{\sigma}$, one usually extracts a complete description
for $\op{\sigma}$ via quantum tomography, and then compares this
description to a theoretical state $\op{\rho}$ by computing the
fidelity $F(\hat\rho,\hat\sigma)$---a single number, commonly used as 
similarity measure. As this example
illustrates, we often have an idea of what has been realized in the
laboratory, so we are interested in asking for much less
information---{\em e.g.}, we only want to know the distance to some
particular theoretical target or to learn the identity of the state or
operation within a restricted set of possibilities.

In this Letter, we develop targeted approaches to directly extract the
information of interest. Our main results, summarized at
Table~\ref{tab:CertificationTable}, show that it is possible to
efficiently characterize a large class of states and
operations---including some that are universal resources for quantum
computation---without resorting to tomography and using only local
measurements and the preparation of product states.  Our methods apply
to discrete variable systems such as qubits, as well as continuous
variable systems such as oscillators. We consider two types of
characterization: {\em certification} and {\em learning}.
  
{\em Learning} consists of identifying the theoretical description
from a restricted set of possibilities that best matches the
experimental data. There exists many classes of ``variational" states
in physics that can be specified with a small number of parameters. 
We provide examples where these parameters can be extracted directly
from experiments, circumventing tomography and hence drastically reducing
the complexity.

{\em Certification} consists of estimating the fidelity between an
experimental device and some theoretical target.  We demonstrate that
certification always requires drastically less resources than full
tomography---in some important cases, it is an exponential reduction
in resources.  Even in the worst case, our scheme offers four
significant advantages for the characterization of quantum states
(equivalent statements hold for quantum operations): {\em (1)} Its
computational cost is bounded by $n^24^n$, compared to $4^{3n}$
required for the simplest tomography procedure based on
pseudo-inverses. {\em (2)} The number of distinct experimental
settings it requires is constant---independent of the system size and
depending only on the desired accuracy of the estimate---compared to
the $4^n$ distinct experiments needed by tomography, or the
$\cO(n2^n)$ settings required by compressed sensing
techniques~\cite{GLF+10}. {\em (3)} The total number of measurements
(counting repeated measurements used to statistically estimate
expectation values) of our scheme is bounded by $\cO(2^n)$, which is
at least a quadratic improvement over what is required by full
tomography. {\em (4)} The data post-processing of our scheme is
trivial, while the correct method of processing tomography data is a
matter of current debates and different methods produce significantly
different results \cite{Blu10}.

The rest of this Letter is structured as follows. In the next three
sections, we describe the state certification scheme for qubits, show
how it extends to continuous variable systems, and the certification
of quantum processes. Then, we present concrete examples drawn from
Table~\ref{tab:CertificationTable}.

\begin{table}
\scriptsize{
\begin{tabular}{|c|c||c|c||c|}
\cline{3-5} 
\multicolumn{1}{c}{} &  & \multicolumn{2}{c||}{\textbf{Certification}} & {\textbf{Learning}}\tabularnewline
\cline{3-4} 
\multicolumn{1}{c}{} &  & Sampling ({\bf C1}) & Fluctuations ({\bf C2}) & \tabularnewline
\hline
States & Stabilizer & \color{red}{$\cO(n)$} &  \color{red}{$\cO(1)$} &   \color{red}{${\mathrm{poly}}(n)$} \tabularnewline
\cline{2-5} 
& W &  \color{red}{$\cO(n)$}  &  \color{red}$\cO(n)$  &   \color{red}{$\cO(n)$} \tabularnewline
\cline{2-5} 
& $\ket{t_n}$ &  \color{red}{$\cO(n)$}  &  \color{red}$\cO(n)$  &   \color{red}{$\cO(n)$} \tabularnewline
\cline{2-5} 
& General MPS &  \color{red}{$\cO(n)$}  & ?  &   \color{red}{$\cO(n)$} \cite{CPFS10a} \tabularnewline
\cline{2-5} 
& General pure state & $\cO(n^2 2^{2n})$ & $\cO(2^{n})$ & $\cO(2^{6n})$ \tabularnewline
\hline 
Processes & Clifford &  \color{red}{$\cO(1)$}  &  \color{red}{$\cO(1)$} &   \color{red}{${\mathrm{poly}}(n)$}\tabularnewline
\cline{2-5} 
& MPS Choi matrix &  \color{red}{$\cO(n)$} & ? &   \color{red}{$\cO(n)$} \tabularnewline
\cline{2-5} 
& General unitary & {$\cO(n^2 2^{4n})$} & $\cO(2^{2n})$ &  {$\cO(2^{12n})$}\tabularnewline
\hline
Evolution & Local Hamiltonian & --- &  --- &  \color{red}{$\cO(n)$} \tabularnewline
\cline{2-5} 
 & Local Lindbladian & --- &  --- &  \color{red}{$\cO(n)$} \tabularnewline
\hline
\end{tabular}
}
\par\caption{ 
Complexity of the characterization of various states and
processes. Entries in red are efficient, {\em i.e.} require resources
that grow at most polynomially with the number of qubits $n$.  
The {\bf Sampling} column gives the complexity of the classical
processing required to sample from the relevance distribution, {\em
  c.f.} {\bf C1}. The {\bf Fluctuations} column gives the number of measurements
required to suppress statistical fluctuations  when evaluating the
fidelity, {\em c.f.} {\bf C2}. The {\bf Learning} column gives the
total number of measurements (including repetitions of the same
measurement setting) required to learn the state within a restricted
set; the classical processing is always a polynomial of that
number. 
When both fidelity estimate and learning are efficient, it is
not necessary to assume that the state belongs to a restricted set as
fidelity testifies of that assumption.  
Stabilizer states, Clifford
gates, Local Hamiltonians and Lindbladians are discussed in the main
text. The W state has often been used as an experimental benchmark,
{\em e.g.} \cite{HHR+05}. The $\ket{t_n}$ state plays a key role in
linear optics quantum computation \cite{KLM01}. Matrix product states
(MPS) accurately describe ground states of 1D quantum systems
\cite{VC06}.  An important example of a process with MPS Choi matrix
is the approximate quantum Fourier transform \cite{BEST96}, key
component of Shor's factoring algorithm.  Question marks indicate open
problems, but they can be no worst than the general states and
operations.
\label{tab:CertificationTable}}
\end{table}

{\em Monte Carlo state certification}---To estimate the fidelity to
some theoretical pure state $\hat\rho$, we use the fidelity
\begin{equation}
F(\op{\rho},\op{\sigma}) = \tr\hat\rho\hat\sigma
=\sum_i {\rho_i \sigma_i\over d} = \sum_i \frac{\rho_i^2} d \frac{\sigma_i}{\rho_i}.
\label{eq:fidelity}
\end{equation}
where $\rho_i = \tr \op\rho \op P_i$, $\sigma_i = \tr \op\sigma \op
P_i$, $d$ is the dimension of the Hilbert space, and $\op P_i$ is
some orthonormal Hermitian operator basis satisfying $\tr \op{P}_i
\op{P}_j=d\delta_{ij}$. For a system composed of $n$ qubits, the $\op
P_i$ could be the $4^n$ Pauli operators obtained by taking tensor
products of the Pauli matrices and the identity. Defining the {\em
  relevance distribution} $\Pr(i)={\rho_i^2\over d}$, we can rewrite
the fidelity as
$F(\op\sigma,\op\rho)=\sum_i \Pr(i){\sigma_i\over\rho_i}$,
where the sum is taken over only the $i$ with $\rho_i\not=0$.  This
expression leads to an experimental procedure to estimate the fidelity
based on Monte Carlo methods as follows: one generates $N$ random
indices $i_1,i_2,\ldots, i_N$ following the relevance distribution
$\Pr(i)$ and estimates $\sigma_{i_k}=\langle \op P_{i_k} \rangle_{\op
  \sigma}$, the experimental expectation value of the observable $\op
P_{i_k}$. With high probability, the fidelity is close to $\frac 1N
\sum_{k=1}^N \frac{\sigma_{i_k}}{\rho_{i_k}}$ with an uncertainty that
decreases as $\frac 1{\sqrt N}$. The total number of distinct
experimental settings is at most $N$, independent of the system size.

\begin{figure}[!t]
\includegraphics[width=.49\textwidth]{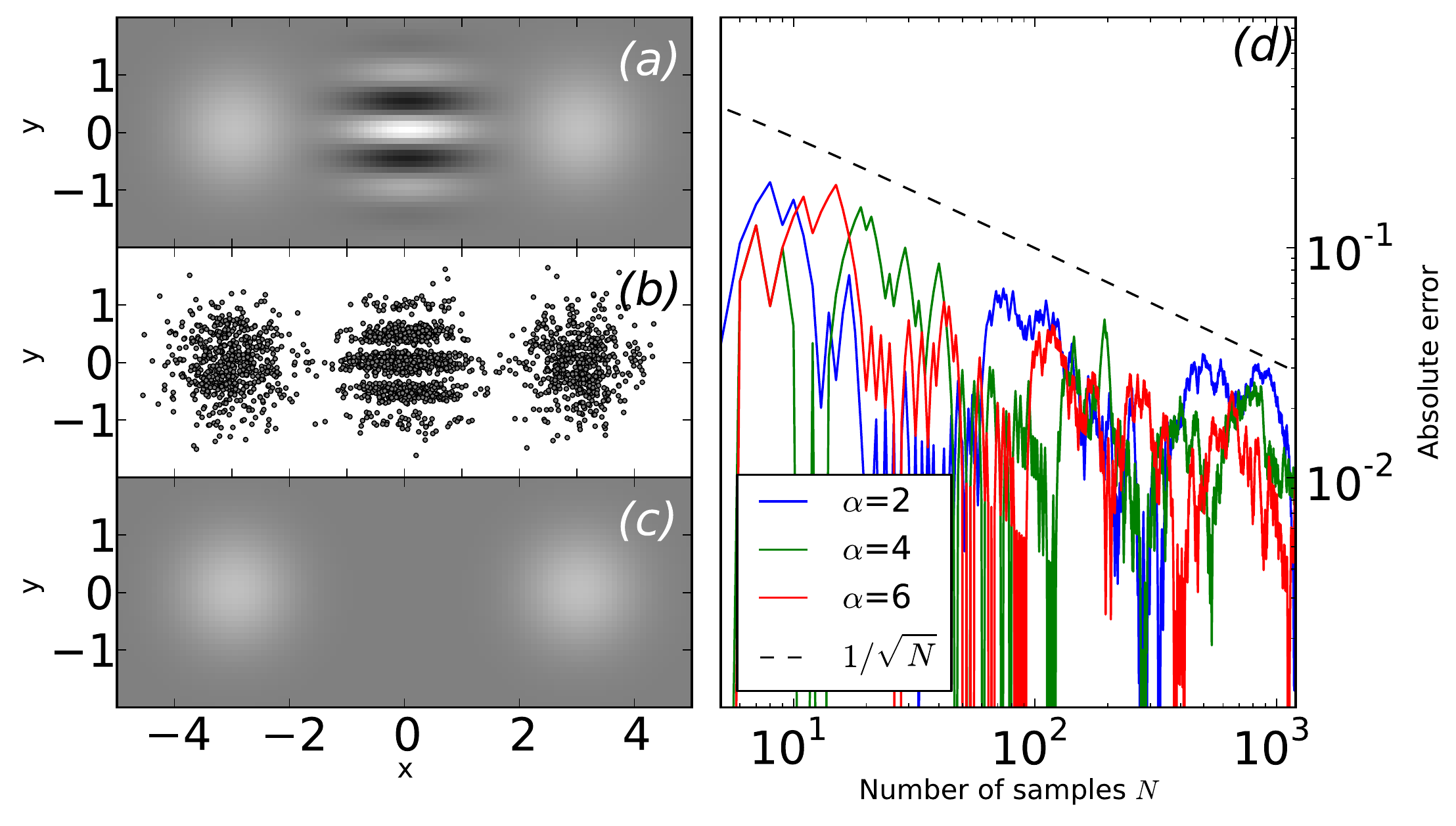}
\caption{(a) Wigner function representation of a harmonic oscillator
  in the superposition $\ket{\psi}=\ket{\alpha}+\ket{-\alpha}$ for
  $\alpha=3$, (b) $10^3$ samples of points in the complex plane drawn
  according to the relevance density of $\ket{\psi}$, (c) Wigner
  function representation of a harmonic oscillator in the incoherent
  mixture of $\ket{\alpha}$ and $\ket{-\alpha}$, corresponding to the
  preparation of $\op\sigma$, (d) absolute error in successive
  estimates of the fidelity $F(\op\rho,\op\sigma)$ for 5 different
  runs with $10^3$ samples each.\label{Wigner}}
\end{figure}

There are two important caveats to this technique:
\begin{enumerate}
\item[\bf C1] Generating an index $i$ according to the relevance
  distribution $\Pr(i)$ can in general require an exponential amount of
  computational resources.
\item[\bf C2] Each $\sigma_{i_k}$ is estimated within some finite
  accuracy. To estimate the fidelity with accuracy $\epsilon$ therefore
  requires repeating the measurement of $P_{i_k}$ roughly $(\epsilon
  \rho_{i_k})^{-2}$ times, which in the worst case grows exponentially
  with the number of qubits.
\end{enumerate}
These are important limitations, and as a consequence our method will
not scale polynomially for all quantum states and operations, but
nevertheless always does significantly better than tomography.  In
addition, there are important classes of states and operations which
avoid these two problems (see Table~\ref{tab:CertificationTable} and
the Supplemental Material for complete details).

\noindent{\em Continuous variables systems---}
For infinite dimensional systems, such as a harmonic oscillator or a
single optical mode in a cavity, it is more convenient to describe a
state $\hat \rho$ by its Wigner functions
$W_{\hat\rho}(\alpha)$~\cite{Wig32} (other indicator functions could
also be used). Equation~\eqref{eq:fidelity} becomes
\begin{equation}
F(\op\rho,\op\sigma) = {1\over\pi}\int_{\mathbb{C}}d^2\alpha~p(\alpha) 
{W_{\op\sigma}(\alpha)\over W_{\op\rho}(\alpha)}
\end{equation}
where the relevance density $p(\alpha) = W^2_{\hat\rho}(\alpha)$ is
defined as the square of the Wigner function of the theoretical state,
whose purity guarantees once again that $p(\alpha)$ is well defined as
a probability density. The Wigner function of the experimental state
$\op\sigma$ can be measured by interactions with an atom and
measurements of the atom's state~\cite{LMK+96}.
Points in the complex plane can be selected according to $p(\alpha)$
using simple methods such as rejection sampling. As an example, we
simulated this proposed method to estimate the fidelity between a
quantum superposition of two harmonic oscillator states---a ``cat''
state $\frac 1{\sqrt{2}}(\ket\alpha +\ket{-\alpha})$---and the
probabilistic mixture of those two classical states. For the given
choice of parameters, this fidelity is $1/2 (1 + e^{-2\alpha^2} )
\approx 0.5$, and Fig.~\ref{Wigner} clearly demonstrates a close
agreement between the Monte Carlo estimate and the exact theoretical
value, as the absolute error decreases like the square-root of the
number of samples of the Wigner function. As expected, the error in
the fidelity estimate does not depend on the state itself
(e.g. average number of photons, amplitude, etc.) but only on the
number of samples.  We emphasize once again that no estimate of the
Wigner function of the experimental state is ever made, so there is no
need for maximum-likelihood fits to the data, or Radon transforms.

{\em Monte Carlo process certification}---The Choi-Jamio{\l{}}kowski
isomorphism~\cite{Jam72} associates to every quantum operation
$\sop E$ on a $d$-dimensional space a density operator $\hat
\rho_{\sop E}$ on a $d^2$-dimensional space via
$
\op{\rho}_{\sop{E}} = \left( \id \otimes \sop{E} \right) ( \ketbra{\phi}\phi)) 
$
where $\ket{\phi}=\frac{1}{\sqrt{d}}\sum_{i=1}^{d}\ket i\otimes\ket i$
and $\id$ is the identity operation.  As with state certification, our
goal is to compare a target unitary $\sop U$ to its experimental
realization $\widetilde{\sop U}$. A good figure of merit in that case is
the average output fidelity $\overline{F}(\sop U,\widetilde{\sop U})$,
defined as the fidelity between the output states produced by $\sop U$
and $\widetilde{\sop U}$, averaged uniformly over all pure input
states. It can be shown that
$
\overline{F}(\sop U,\widetilde{\sop U})=\frac{d~F(\op{\rho}_{\sop
U},\op{\rho}_{\sop{\widetilde{U}}})+1}{d+1}
$~\cite{HHH99},
reducing the problem of comparing two processes $\sop U$ and
$\sop{\widetilde U}$ to the problem of comparing two states
$\op{\rho}_{\sop U}$ and $\op{\rho}_{\sop{\widetilde U}}$. This problem is
solved by the Monte Carlo state certification presented above.

While this derivation makes use of the maximally entangled state
$\ket\phi$, the experimental realization of the protocol requires only
the preparation of product states. A direct implementation of the
quantum Monte Carlo state certification would prepare a maximally
entangled state $\ket\phi$, apply $\sop{\widetilde{U}}$ to half of the
system, and then measure random Pauli operators on all qubits. A more
practical approach consists of preparing the complex conjugate of
random product of eigenstates of local Pauli operators (corresponding
to the resulting state after half of the entangled state is measured
destructively), applying the transformation $\sop{\widetilde{U}}$ to the
system, and finally measuring a random Pauli operator on each
qubit. This simplification, based on the identity
$(\ketbra{\mu}{\mu}\otimes \id) \ket\phi = \ket\mu \otimes \ket{\mu}^*$,
generates the same statistics as the direct scheme~\cite{BPP08}.  

{\em Computation via teleportation}---Some of the most promising
approaches to universal and scalable quantum computation are
teleportation-based quantum computation~\cite{GC99} and
measurement-based quantum computation~\cite{RB01}. Both these
approaches rely heavily on the preparation of stabilizer
states~\cite{Got97} and the application of quantum operations known as
the Clifford group~\cite{GC99}, which map stabilizer states to
stabilizer states.  Stabilizer states are also important for quantum
computation in general because of their close relationship to a large
class of quantum error correction codes known as {\em stabilizer
  codes}. Many of the experimental demonstrations of state preparation
to date have been of stabilizer states, such as states encoded into
stabilizer codes~\cite{CLS+04}, cluster states~\cite{WRR+05}, and the
GHZ state $\ket{00\cdots0}+\ket{11\cdots1}$~\cite{LKS+05,MSB+11}.

We first describe how to {\em certify} these states and operations.
Stabilizer states are defined to be $+1$ eigenstates of some set of
commuting Pauli operators $\op S_j$ that generate the stabilizer
group, {\em i.e.} $\op S_j\ket\psi = \ket\psi$ for all $j=1,\ldots
n$. It follows that $\Pr(i)=1/d$ if either of $\pm\op P_i$ is in the
stabilizer group and $0$ otherwise. Sampling from $\Pr(i)$ thus
amounts to generating an index $i$ uniformly between 1 and $d$,
avoiding the problem associated with caveat {\bf C1}. For the same
reasons, $\rho_i^2 = 1$ for all $i$ with $\Pr(i) \neq 0$, so that the
uncertainty in the estimation of $\sigma_i$ is not amplified,
avoiding the problem associated with caveat {\bf C2}. It also follows
that the fidelity $F(\op\sigma,\op\rho)$ to a stabilizer state
$\op\rho$ can be estimated with error $\epsilon$ using
$N=\cO({1\over\epsilon^4})$ experiments involving only local
projective measurements, {\em independently of the system size and
  without any prior knowledge of the experimental state
  $\op{\sigma}$}.  Since this result relies only on local
measurements, it can immediately be generalized to states which are
locally equivalent to stabilizer states.

This result carries over directly to the certification of Clifford
operation because their Choi-Jamio{\l{}}kowski density operators are
stabilizer states.  In the case of Clifford transformations similar
results can be obtained using ``twirling''
experiments~\cite{ESM+07} or by the selective measurement of
matrix elements of the Choi matrix~\cite{BPP08}, although the Monte
Carlo approach described here generalizes to other cases.

While operations in the Clifford group are not sufficient to perform
universal computation~\cite{GC99}, single qubit rotations can be used
to reach universality, and these can be certified efficiently thanks
to local equivalence of either operations (if the rotation is applied
directly) or state preparation (if the rotation is applied via ``magic
state'' teleportation~\cite{GC99,BK05}).

Stabilizer states can also be {\em learned} efficiently, as pointed
out by Aaronson and Gottesman~\cite{Got08}, although the known method
for efficient stabilizer learning requires entangling
measurements. Aside from the direct generalizatin of the stabilizer
approach, Clifford group operations can be learned
efficiently~\cite{Low09} if one has access to Bell measurements and
the inverse of the operation being learned. The problem of performing
these tasks efficiently with strictly local measurements and without
the need for the inverse remains open.

{\em Local Hamiltonians and Lindbladians}---Models of universal
quantum computation exist where the idea of discrete gates is not a
natural fit. Instead, the system evolves in a continuous way, governed
by some dynamical equation $\frac \partial{\partial t} \op\rho =
\mathcal{G} \op\rho$.  The most direct way to determine how accurately
these dynamics can be realize is to estimate the time evolution
generator $\mathcal G$ of the system, and explicitly check how it
compares against the ideal target generator. Important examples
include local Hamiltonians and Lindbladians that are universal for
adiabatic quantum computation \cite{ADK+07} and dissipation-driven
quantum computation \cite{VWC09} respectively.

In what follows we demonstrate how to learn such local $\mathcal G$ using
only {\em (i)} the preparation of initial product states, {\em (ii)}
the simultaneous measurement of a constant number of single-qubit
operator, {\em (iii)} a number of experimental settings that grows
linearly with the system size, {\em (iv)} and classical
post-processing of complexity $n^3$ (inverting an $cn\times cn$ matrix
for some constant $c$); improving on~\cite{SML+10}.

Consider the case of coherent evolution generated by some Hamiltonian
$H$. For a short time $t$, the expectation value of any observable
$\hat A$ evolves as
\begin{equation}
\langle \hat A(t)\rangle_{\hat
  \rho} - \tr \hat A\hat \rho = it\langle[\hat H,\hat A]\rangle_{\hat
  \rho} + \cO(\|\hat H\|^2 t^2).
  \label{eq:linear_dynamics}
\end{equation}
 By experimentally measuring this
expectation value, we obtain one linear constraint on the Hamiltonian.
Varying over different observables $\hat A_i$ and initial states $\hat
\rho_j$, we obtain more linear constraints that we can write as
$W_{ij} = \langle \hat A_i(t)\rangle_{\hat \rho_j} - \tr \hat A_i\hat
\rho_j = it\langle[\hat H,\hat A_i]\rangle_{\hat \rho_j}$ where we
have dropped the higher order terms $\cO(\|\hat H\|^2 t^2)$. Writing
$\hat H$ in an operator basis $\hat H = \sum_l h_l \hat P_l$, we
obtain the linear equation 
$W_{ij} = \sum_l T_{ij,l} h_l \label{eq:LI}$ where
$T_{ij,l} = it \tr \hat \rho_j [\hat P_l,\hat A_i]$. The Hamiltonian
can be learned by inverting this linear equation~\cite{SML+10}.
 
There are in general a number important caveats to this approach,
although all of these disappear when the Hamiltonian is {\em local},
which is nonetheless sufficient to achieve universal quantum
computation \cite{ADK+07,VWC09}. The Lieb-Robinson bound \cite{LR72a}
shows that only the Hamiltonian $\hat H_R$ in a region $R$ a distance
$d\approx vt$ of the local observable $\hat A$ contributes to its
evolution, {\it i.e.}, $e^{i\hat H t}\hat A e^{-i\hat H t} \approx
e^{i\hat H_R t}\hat A e^{-i\hat H_R t}$ (for details of the proof see
the Supplemental Material). This fact solves all the problems
associated to the proposal of \cite{SML+10}:

{\em 1)} The error $\cO(\|H\|^2 t^2)$ appearing in
Eq.~\eqref{eq:linear_dynamics} becomes $\cO(\|H_R\|^2 t^2) = \cO(\|\hat
A\|^2 t^4)$, independent of the system size. Thus, it is not necessary
to decrease the evolution time $t$ as the system size increases to
achieve a given accuracy.

{\em 2)} Because the Hamiltonian is local, the number of non-zero
terms $h_l$ is proportional to the number of particles in any finite
dimension. Thus, in the linear equation for $W_{ij}$, the range of the
index $l$ increases only linearly with the number of particles, as
opposed to the exponential growth for generic Hamiltonians.

{\em 3)} Because the dynamics is local, $T_{ij,l} = T_{ij',l}$ when
$\hat \rho_j$ and $\hat \rho_{j'}$ differ only outside a region of
radius $k$ away from the local observable $\hat A_i$. In addition, the
$T$ become linearly dependent---and thus redundant---when the input
states are linearly dependent.  For each observable $\hat A_i$,
we only need to vary the initial state locally, so the total number of
observable-state pairs $(ij)$ grows linearly with the number of
particles. Thus, learning the Hamiltonian---or equivalently the
$h_l$---amounts to inverting the linear-size linear equation $W_{ij} =
\sum_l T_{ij,l} h_l$.

{\em 4)} Product input states form a complete operator basis, so they
are sufficient to gain all information about the Hamiltonian. Thus
$\tr \hat A_i\hat \rho_j$ can be easily computed since $\hat A_i$ is
local and $\hat \rho_j$ is a product state. The quantity $\tr \hat
\rho_j [\hat P_l,\hat A_i]$ can also be evaluated efficiently because
the commutator of two $k$-local operators is at most $2k$-local, and
$\hat \rho_j$ is a product state.

{\em Acknowledgments}---
We thank P. S{\'e}mon for many instructive discussions about Monte
Carlo methods. After this work was made public, some of our findings
were independently derived by Flammia and Liu~\cite{FL}. This
work is partially funded by FQRNT, NSERC, and numerical calculations
were performed using resources from RQCHP.

\appendix

\section{Statistical bound for Monte-Carlo estimation of the fidelity}

We present rigorous bounds for the error of the Monte Carlo fidelity
estimate in the case of an $n$ qubit system. The result ({\it c.f.}
Eq.~\eqref{eq:stat-bound}) can be adapted to the case of continuous variable
systems through the minor modification of replacing the expectation 
value $\rho_{i_k}$ by the value ${1\over2}W_{\op\rho}(\alpha_{i_k})$ of the
Wigner function of state $\op{\rho}$ at point $\alpha_{i_k}$.

\begin{thm}
  Let $\op{\rho}=\sum_{i}\frac{\rho_{i}}{d}\op{P}_{i}$ be the
  decomposition of the pure state $\op{\rho}$ over the orthogonal
  Hermitian operator basis $\{\op{P_{i}}\}$ where
  $\tr\op{P_{i}}\op{P_{j}}=d\delta_{ij}$ and the operator norm $\|\op
  P_i\|\le1$.  One can obtain an estimate $\bar F$ of the fidelity
  $F(\op{\rho},\op{\sigma})$ between $\op{\rho}$ and $\op{\sigma}$
  with error $\epsilon=\epsilon_{1}+\epsilon_{2}$ such
  that
\begin{multline}
\Pr(|F-\bar{F}|\geq\epsilon)
\leq
\frac{1}{N_{1}\epsilon_{1}^{2}}+\\
2\exp\left[-{\epsilon_{2}^{2}N_1^2\over 2}\left(\sum_{k=1}^{N_1}{1\over
\rho_{i_k}^2 N_2^{[k]}}\right)^{-1}\right]
\label{eq:stat-bound}
\end{multline}
  where
\begin{itemize}
\item $I=\{i_{1}\dots i_{N_{1}}\}$ are $N_1$ indices sampled from $\Pr(i)$,
corresponding to observables $\hat P_{i_k}$ to be measured experimentally on
$\op{\sigma}$
\item $N_{2}^{[k]}$ is the number of experimental samples taken to estimate
$\sigma_{i_{k}}=\tr\op{P}_{i_k}\op{\sigma}$
\item $\epsilon_{1}$ is the error associated to the Monte Carlo estimate
\item $\epsilon_{2}$ is the error associated to the experimental estimation
of the $\{\sigma_{i}\}_{i\in I}$
\end{itemize}
\end{thm}
\begin{proof}
  The fidelity $F(\op{\rho},\op{\sigma})$ can be rewritten
  as 
\begin{equation}   
F(\op{\rho},\op{\sigma})=\sideset{}{'}\sum_{i}\frac{\rho_{i}^{2}}{d}\frac{
\sigma_{i}}{\rho_{i}}
\end{equation}
  where prime indicates that the summation runs only over non-zero values
  of $\rho_{i}$. Since $\tr\op{\rho}^{2}=1$ by assumption,
  $\Pr(i)=\rho_{i}^{2}/d$ is a normalized probability distribution.
  We can thus interpret the fidelity as the expectation value of a
  random variable $X$ which takes value $\sigma_{i}/\rho_{i}$ with
  probability $\Pr(i)$. Its variance is bounded by a constant,
  as 
\begin{equation}
    \mbox{Var}(X)=\sideset{}{'}\sum_{i}{\sigma_{i}^{2}\over
d}-F^{2}\leq\tr\op{\sigma}^{2}-F^{2}\leq1,
\end{equation}
  and thus, using Chebyshev's inequality, we obtain
\begin{equation}
    \Pr(|F-\bar{F}_{1}|\geq\epsilon_{1})\leq\frac{1}{N_{1}\epsilon_{1}^{2}},
\end{equation}
  where $\bar{F}_{1}=\sum_{i\in I}\sigma_{i}/\rho_{i}$ is the estimate
  of the fidelity by sampling $N_{1}$ realizations of $X$, {\em i.e.}, by
  drawing $I=\{i_{1}\dots i_{N_{1}}\}$ indexes from the probability
  distribution $\Pr(i)$ and estimating $\mathbb{E}(X)$ by the
  realization of $\bar{X}=N_{1}^{-1}\sum_{i\in I}X_{i}$ where all
  $X_{i}$ are independent and distributed as $X$. Thus, the number of
measurements
  settings does not depend on the dimension of the system and scales
  as $\cO(1/\epsilon_{1}^{2})$.

  The expectation value $\sigma_{i}$ of each observables with
  respect to the experimental state $\op{\sigma}$ can only be
  estimated up to finite precision. For each $i_k \in I$, the observable
  $\op{P}_{i_{k}}$ is measured on the experimental state and yields a
  number $y^{[m]}_{i_{k}}$ whose absolute value is bounded by the
  operator norm of the observables. This measurement is repeated
  $N_{2}^{[k]}$ times and the approximate realization of $X_{k}$ is
$\tilde{\sigma}_{i_{k}}/\rho_{i_{k}}=\left(\rho_{i_{k}}N_{2}^{[k]}\right)^{-1}
\sum_{m=1}^{N_{2}^{[k]}}y^{[m]}_{i_{k}}$.
  This estimation proceadure is then repeated for each of the $N_{1}$
  observables. Hoeffding's bound~\cite{Hoe63} states that, if the
  \emph{independent} real random variables $Y_{i}$ are such that $a_{i}\leq
  Y_{i}\leq b_{i}$, then for $S=Y_{1}+Y_{2}+\cdots+Y_{n}$,
\begin{equation}
\Pr(|S-\langle S\rangle|\ge
t)\le2\exp\left(-\frac{2t^{2}}{\sum_{i=1}^{n}(b_{i}-a_{i})^{2}}\right) \mbox{.}
\end{equation}
In our case, for all $k$, we have
$-1 / \left|\rho_{i_{k}}\right| \leq y_{i_{k}}^{[k]} / \rho_{i_{k
} }  \leq 1 /  \left| \rho_{i_{k}}\right|$
for all $N_{2}^{[k]}$ experimental measurement performed to estimate
$\sigma_{i_{k}}$ and we can apply the Hoeffding's inequality to all
$N=\sum_{k=1}^{N_{1}}N_{2}^{[k]}$ experimental samples to bound the
distance between the sum $\bar{F}$ of $\tilde{\sigma}_{i_{k}}/\rho_{i_{k}}$
by 
\begin{equation}
\Pr(|\bar{F}-\bar{F}_{1}|\geq\epsilon_{2})\leq
2\exp\left[-{\epsilon_{2}^{2}N_1^2\over 2}\left(\sum_{k=1}^{N_1}{1\over
\rho_{i_k}^2 N_2^{[k]}}\right)^{-1}\right].
\label{eq:Hoeffding_allsamples}
\end{equation}
Finally to reach eq. \eqref{eq:stat-bound}, one applies the union
bound to $|F-\bar{F}|\leq|F-\bar{F}_{1}|+|\bar{F}-\bar{F}_{1}|$.
\end{proof}

As can be seen in the last term of Eq.~\eqref{eq:stat-bound},
observable $\op P_{i_k}$ must be sampled $N_2^{[k]} \gg \rho^2_{i_k}$
times to obtain an accurate estimate of its expectation value. While
this can be large in general, there are many important cases where the
$\rho_i$ are only polynomially small, leading to a polynomial
$N_2^{[k]}$. Two examples of cases of interest beyond the stabilizer states
presented in main text are the $W$ state~\cite{HHR+05} and the $\ket{t_n}$ state
used in linear optics for heralded
teleportation with high success probability ~\cite{KLM01}. Both are
MPS with bond dimension 2 and are uniform superpositions of a linear
number of computational-basis states.  For both states, the
expectation value of a Pauli operator $\op P$ is given by
\begin{equation}
 \langle \psi | \op P | \psi \rangle = \alpha(n) \sum_{i,j} \langle i | \op P |
j\rangle
\end{equation}
where $\alpha(n)$ is $1/n$ for the W state and $1/(n+1)$ for
$\ket{t_n}$, and the sum runs over
computational states that appear in the decomposition of the
state. For all $i,j$, there exists a Pauli operator $\op{\sigma}_{ij}$
such that $\ket j = \op{ \sigma}_{ij} \ket i$. Since the Pauli
operators form a group, $\op P \op {\sigma}_{ij} $ is another Pauli
operator and all terms appearing in the sums are $\pm 1$. Thus, the
smallest non-zero Pauli expectation scales as $1/n$, and the number of
samples required to estimate $\sigma_i/\rho_i$ to constant accuracy
scales as $n^2$ in the worst case.

More generally, we can improve the error bound
Eq.~\eqref{eq:stat-bound} by truncating the relevance
distribution. Define the set of negligible expectation values as
$S\equiv\left\{ \rho_{i}\mbox{ such that
  }|\rho_{i}|<d^{-\alpha}\right\} $ where $\alpha$ is a positive
number to be determined.  We split the fidelity into a significant and a
negligible contribution
\begin{equation}
  F(\op \rho,\op\sigma) = \sum_{i}\frac{\rho_{i}\sigma_{i}}{d}=\sum_{\rho_{i}\notin
    S}\frac{\rho_{i}\sigma_{i}}{d}+\sum_{\rho_{i}\in
    S}\frac{\rho_{i}\sigma_{i}}{d}
\end{equation} 
and bound the negligible contribution using
\begin{equation}
\left|\sum_{\rho_{i}\in
S}\frac{\rho_{i}\sigma_{i}}{d}\right|\leq\sum_{\rho_{i}\in
S}\frac{\left|\sigma_{i}\right|}{d}\max_{i\in S}\left|\rho_{i}\right|\leq
d^{-(\alpha+1)}\sum_{\rho_{i}\in
S}|\sigma_{i}|\mbox{.}\label{eq:bound_on_sum}
\end{equation}
The sum of a subset of $\left|\sigma_{i}\right|$ is bounded by the sum
over \emph{all} $\left|\sigma_{i}\right|$.  To bound
$\sum_{i}|\sigma_{i}|$, we can use the constraint on the purity of the
state $\sum_{i}\sigma_{i}^{2}=d\,\tr\hat{\sigma}^{2}\leq d$.  The sum
of absolute values is maximal when all absolute values are equal,
which follows from standard Lagrange multiplier techniques. The purity
constraint finally leads to 
\begin{equation} \sum_{i}|\sigma_{i}|\leq
  d\sqrt{d\,\tr\hat{\sigma}^{2}}\leq
  d^{3/2}\mbox{.}\label{eq:lagrange}
\end{equation}
Inserting this inequality that into eq. \eqref{eq:bound_on_sum} yields
\begin{equation}
\left|\sum_{\rho_{i}\in S}\frac{\rho_{i}\sigma_{i}}{d}\right|\leq
d^{1/2-\alpha}\mbox{.}
\end{equation}
Hence, the sum over negligible $\rho_{i}$ vanishes exponentially for
$\alpha=(1+\epsilon)/2$, \emph{i.e.}, when we drop all expectation
values smaller than $d^{-\frac{1+\epsilon}2}$ in absolute value, for
any constant $\epsilon>0$.

We thus modify the sampling method in the following way. For each
observable $\hat{P}_{i}$ picked from sampling the relevance
distribution, compute the corresponding expectation value
$\rho_{i}=\tr\hat{\rho}\hat{P}_{i}$. When
$\rho_{i}{}^{2}<d^{-1-\epsilon}$, reject this entry, otherwise you
proceed as before.  It is important to verify that this modification
does not slow down the procedure, \emph{i.e.} that we are not constantly
rejecting samples. To see this, notice that the probability of
choosing an element from the negligible set is bounded by
\begin{equation}
\sum_{\rho_{i}\in S}\frac{\rho_{i}^{2}}{d}\leq\sum_{\rho_{i}\in
S}d^{-2-\epsilon}\leq d^{-\epsilon}\mbox{.}
\end{equation}

Since we reject all negligible $\rho_{i}$, the maximum number of
repeated measurements needed for a given experimental setting scales
in the worst case as $d^{1+\epsilon}$. In particular, for qubits, the
maximum number of measurements is $2^{n(1+\epsilon)}$. Moreover, since
the number of measurement settings does not scale with the size of the
system, the total number of measurements scales as $\cO(2^{n(1+\epsilon)})$
which is at least a \emph{quadratic improvement over the number of measurements
  needed to perform brute-force tomography} on a generic state of $n$
qubits.

\subsection*{Extension to continuous variables systems}

The Monte Carlo
method proposed here can be adapted to continuous variable systems,
such as a single electromagnetic field mode in a cavity \cite{RBH01,
  BK98,HWA+09}, by modifying how the state is parameterized and how
the sampling is performed. The main reason for this is the obvious
difficulty of measuring observables in a discrete infinite dimensional
operator basis. This problem can be avoided by considering phase-space
quasiprobability distribution descriptions of quantum states. If we
consider the dual phase-space distributions $f_{\op{\rho}}(\alpha)$
and $g_{\op{P}_i}(\alpha)$ which correspond respectively to the
quantum state $\op{\rho}$ and an observable $\op{P}_i$~\cite{CG69},
then
$
\tr \op{\rho}\op{P}_i = {1\over\pi} \int_{\mathbb{C}} d^2\alpha~
f_{\op{\rho}}(\alpha) g_{\op{P}_i}(\alpha).
$
It follows that the fidelity between a pure state $\op\rho$ and an
arbitrary state $\op\sigma$ is given by
$
F(\op\rho,\op\sigma) = \tr \op\rho~\op\sigma
= {1\over\pi}\int_{\mathbb{C}}~d^2\alpha~f_{\op{\rho}}(\alpha)~g_{\op\sigma}(\alpha),
$
which can be re-written as
$
F(\op\rho,\op\sigma) 
= {1\over\pi}\int_{\mathbb{C}} d^2\alpha
~p(\alpha)~{g_{\op\sigma}(\alpha) \over f_{\op{\rho}}(\alpha)},
$
where the integration excludes regions with $f_{\op{\rho}}(\alpha)=0$
and where $p(\alpha) = f^2_{\op{\rho}}(\alpha)$ is the {\em relevance
  density function}. Sampling the relevance density can be done by
standard methods, such as rejection sampling.

The choice of phase space distributions is important, as it must be
possible to interpret $f^2_{\op{\rho}}(\alpha)$ as probability
distributions, and it must be possible to estimate
$g_{\op\sigma}(\alpha)$ at some arbitrary $\alpha\in{\mathbb{C}}$
easily from experimental data. One choice that fulfills both these
requirements for all states is the Wigner
function~\cite{Wig32,CG69}. The Wigner function is self-dual and
bounded in magnitude by $2$, and its value at particular $\alpha$ can
be estimated by using simple experiments where the continuous variable
system, such as an electromagnetic field mode, interacts with an
atom~\cite{LMK+96,LD97,BAM+02,HWA+09}.

The same truncation technique used to evaluate the performance of this
algorithm for qubits can be used for continuous variable systems.
Amplification of experimental uncertainty can once again by reduced by
placing a cut-off in the relevance density function. If we disregard
regions in phase space where the absolute value of the relevance
density is below $c$, then the error $E$ in the fidelity is bounded by
\begin{align}
E 
& = {1\over\pi}\left| \int_I d^2\alpha~W_{\op\rho}(\alpha) 
W_{\op\sigma}(\alpha)\right|,\\
& \le  {1\over\pi}\sqrt{\int_I d^2\alpha~W^2_{\op\rho}(\alpha)}
\end{align}
where $I$ is the region in phase space where $|W_{\op\rho}|<c$.

\section{Sampling from the relevance distribution}

Sampling from the relevance distribution $\Pr(i)$ is not trivial
because the dimension of the operator space on $n$ particles is exponentially
large in $n$. Therefore, computing all $\rho_{i}=\tr\hat{\rho}\hat{P}_{i}$
for all observables $\hat{P}_{i}$ is unefficient. Furthermore, computing
a given $\rho_{i}$ can be a challenging task in itself. However, by choosing
operators $\hat P_i = \hat p_i^{[1]} \otimes \ldots \otimes \hat p_i^{[n]}$ that
are tensor products of single-particle operators---such as the Pauli operators
for qubits---sampling can be
simplified by recursively picking the observables for each particle as we now
demonstrate.

\subsection{Sampling using conditional probabilities}

Consider for concreteness a system composed of $n$ qubits, and an
operator basis $\hat P_i$ all consisting of tensor product of single
qubit operators, {\it e.g.} Pauli operators. The Hilbert space
dimension is $d = 2^n$.  For an observable
$\hat{P}_{i}=\bigotimes_{m=1}^{n}\hat{p}_{i_{m}}^{[m]}$, denote the
relevance distribution $\Pr(i) = q_{i_1,\dots,i_n}$.  Using the
probability chain rule, this probability can be expressed as a product
of conditional probabilities
\begin{equation} 
q_{i_1,\dots,i_n}=\prod_{k=1}^{n}
  q_{i_k \left| i_1,\dots,i_{k-1} \right. }
\label{eq:Marginal}
\end{equation}
where the conditional probability $q_{i_k \left| i_1,\dots,i_{k-1}
  \right. }$ of drawing the observable $\hat{p}{}_{i_{k}}^{[k]}$ on
particle $k$ knowing which observables have been picked on the
previous particles is
\begin{equation}
q_{i_k \left| i_1,\dots,i_{k-1} \right. }
= q_{ i_1,\dots,i_{k-1}  }^{-1} 
\sum_{I=i_{k+1},\dots,i_n} q_{ i_1,\dots,i_{k} I }\mbox{.}
\label{eq:MarginalDistribution}
\end{equation}
Using equation \eqref{eq:Marginal}, sampling from the probability
distribution reduces to sequentially picking an observable
$\hat{p}_{i_{m}}^{[m]}$ according to the conditional probability
distribution~\eqref{eq:MarginalDistribution} which can be written, up
to a normalization factor, as
\begin{align*}
  q_{i_k \left| i_1,\dots,i_{k-1} \right. }
  & \propto 
  \sum_{\hat{P}\in\mathcal{P}_{n-k}}\tr\left[\left(\hat{\rho}
      \times\left(\bigotimes_{m=1}^{k}\hat{p}_{i_{m}}^{[m]}\otimes\hat{P}
      \right)\right)^ { \otimes2 }
  \right],\label{eq:2copies} \\
  &= \tr\left[\hat{\rho}^{\otimes2}\left(\bigotimes_{m=1}^{k}\left(\hat{p
        }_{i_{m}}\right)^{\otimes2}\otimes\sum_{\hat{P}\in\mathcal{P}_{n-k}}\hat{P}^{
        \otimes2}\right)\right]
\end{align*}
where the trace of two copies accounts for the square in the
definition of $\Pr(i) = \frac{\tr (\op\rho \op P_i)^2}{d} =  \frac{\tr (\op\rho
\otimes \op \rho \op P_i\otimes \op P_i)}{d}$.  The sum
over all duplicated observables $\hat{P}^{\otimes2}$ can be written as
the tensor product of operators acting on each pair $[m,\, n+m]$ of
particles\begin{equation}
  2^{-(n-k)}\sum_{\op{P}\in\mathcal{P}_{n-k}}\op{P}\otimes
  \op{P}=\bigotimes_{m=k+1}^{n}\op{\Omega}^{[m,\, n+m]}\end{equation} where
$\op{\Omega}^{[i,j]}=\frac{1}{2}\sum_{m}\hat{p}_{m}^{[i]}\otimes
\hat{p}_{m}^{[j]}$ is an observable acting on the pair of particles
$(i,j)$. For instance, for the Pauli operator basis, $\op{\Omega}$ is
the SWAP operator. Thus, the conditional probability is proportionnal
to
\begin{equation}
\tr\left[\hat{\rho}^{\otimes2}\left(\bigotimes_{m=1}^{k}\left(\hat{p
}_{i_{m}}\right)^{\otimes2}\bigotimes_{m=k+1}^{n}\op{\Omega}^{[m,\,
n+m]}\right)\right]\label{eq:MarginalDistTwoCopies}\end{equation}
which is the \emph{expectation value of a tensor product of 2-local
observables} on the state $\hat{\rho}\otimes\hat{\rho}$ on $2n$ particles.

\subsection{Bound on the complexity of sampling}

The problem of sampling reduces to, for each of the $n$ particles,
{\em i)} computing conditional probabilities for each of the possible
observables acting on that particle {\em ii)} pick one of those
observables by generating a random number. Conditional probabilities
can be expressed as expectation values through
eq.~\eqref{eq:MarginalDistTwoCopies}. Thus, if computing expectation
values on tensor product of local observables on states of $n$
particles has complexity $q(n)$, generating an index $i=i_{1}\dots
i_{n}$ from the relevance distribution Pr($i$) has complexity at most
$n\times q(2n)$.

\begin{figure}[b]
\includegraphics[width=0.9\columnwidth]{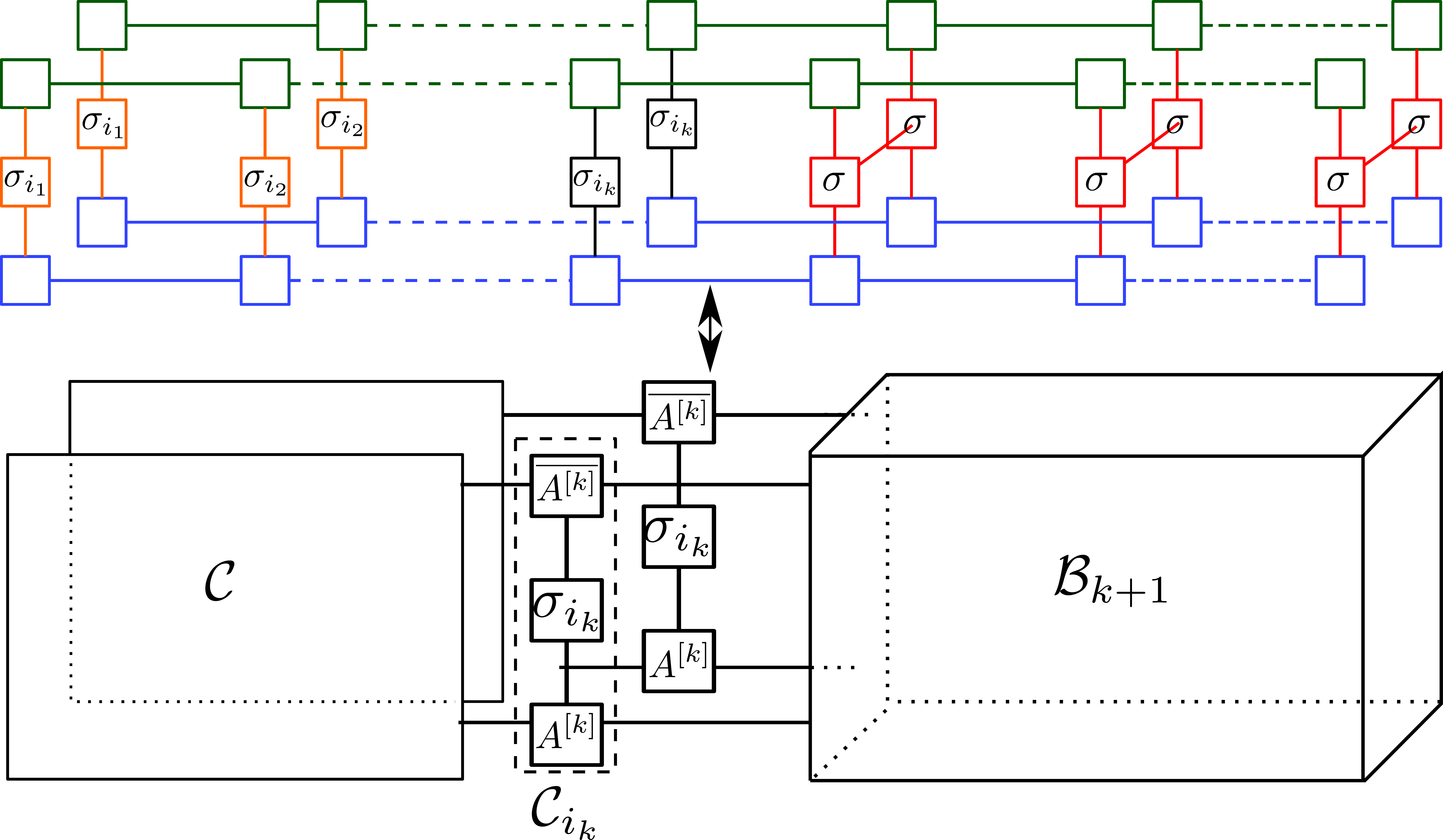}
\caption{\label{fig:MPSsampling} Tensor network corresponding to
eq. \eqref{eq:MarginalDistTwoCopies} if $\hat{\rho}=|\psi\rangle\langle\psi|$ is
a MPS, \emph{i.e.}, there exist a familty of matrices
$\left\{A^{[k]}_{i_k}\right\}$
such that $\ket{\psi}=A_{i_{1}}^{[1]}\dots A_{i_{n}}^{[n]}\ket{i_{1}\dots
i_{n}}$. The upper figure represent the individual tensors in the tensor
network. Each square represent a tensor and outgoing legs represent the tensor
indices. Two squares connected by a line are the contraction of the
corresponding indices of two tensors. Red squares correspond to the $\Omega$
operators. Orange squares correspond to the Pauli operators already chosen on
the $k-1$ previous qubits. Blue squares are the MPS tensors of the two copies of
$\ket{\psi}$ while the green squares are the MPS tensors of the two copies of
$\bra{\psi}$. The lower figure correspond to the partial contraction of
the tensor network.}
\end{figure}

For many states of interest, computing expectation values of local
observables can be performed in polynomial time, \emph{i.e.}, $q(n)
\in \mbox{poly}(n)$.  That is the case for many families of
tensor-network states such as matrix product states (MPS)~\cite{AKLT}
which are known to represent faithfully ground states of interesting
many-body Hamiltonians in 1D~\cite{VC06}.  In fact, the procedure
outlined above can be simplified in the case of MPS, yielding a
sampling complexity {\em linear} in $n$, see Fig.~\ref{fig:MPSsampling}.  Their natural extension to
2D, projected entangled pair states (PEPS)~\cite{PEPS} also allows the
efficient heuristic computation of such expectation values.

A larger class of multi-qubit states for which sampling can be done
efficiently by computing conditional probabilities are computationally
tractable (CT) states~\cite{Nes09}.  CT states are states in which
{\em (a)} the overlap with any element of the computational basis can
be computed efficiently, and {\em (b)} it is possible to sample from
the distribution of outcomes from measurements in the computational
basis efficiently. For such states, it is possible to efficiently
compute the expectation value of tensor products of Pauli observables
which only permute elements of the computational basis and thus are
basis preserving.

In the generic case of a state defined as a vector of the Hilbert
space, computing the expectation value of a single local observable
will take time $\mathcal{O}(2^{2n})$ since we have to account for the
Hilbert space of $2n$ qubits. A tensor product of local observables
can be thought as the product of $\mathcal{O}(n)$ observables that act
non-trivially on a few qubits. Thus, computing the expectation value
given by equation \eqref{eq:MarginalDistTwoCopies} will take time
$\mathcal{O}\left(n\,2^{2n}\right)$. In order to sample, such a
computation has to be repeated for each particles, leading to an
overall complexity of sampling from the relevance distribution of
$\mathcal{O}\left(n^2\, 2^{2n}\right)$ in the worst case. Learning
algorithms based on compressed sensing can recover low-rank density
matrices from $\mathcal{O}\left(n\, 2^{n}\right)$ expectation values
in any basis~\cite{GLF+10}, which indicates that it may be possible to
improve the performance of the algorithm proposed here in the case of
general pure states.

\section{Lieb-Robinson bound}

\begin{figure}
\includegraphics[width=.2\textwidth]{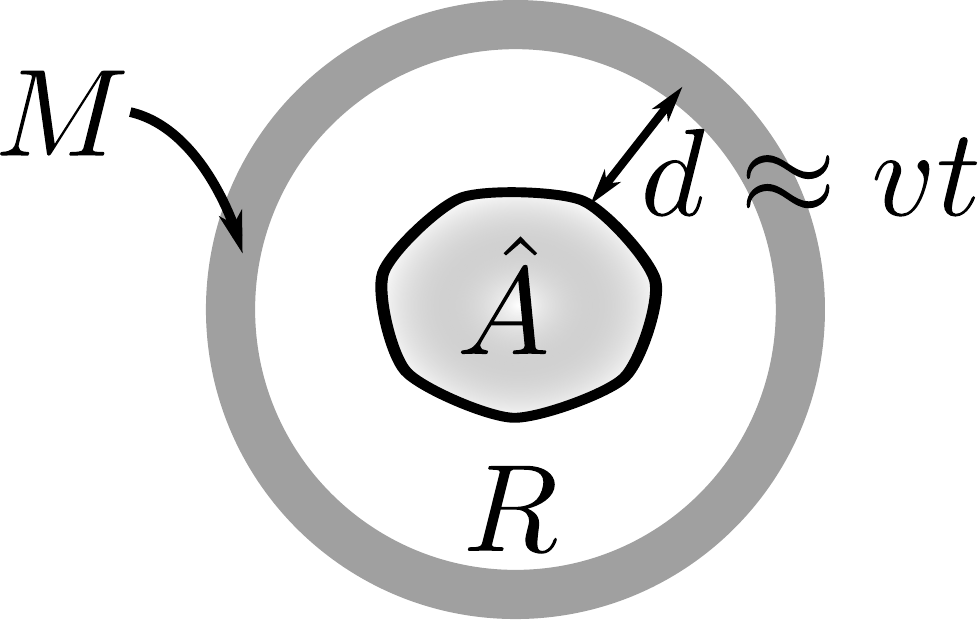}
\caption{
When the system evolves under a {\em local} Hamiltonian (or
Lindbladian), the operator $A$ evolves under the full Hamiltonian $H$
for a time $t$ is essentially the same as the operator resulting from
the evolution generated by the Hamiltonian truncated to the region
$R$. Mathematically, $e^{i\hat H t}\hat A e^{-i\hat H t} \approx
e^{i\hat H_R t}\hat A e^{-i\hat H_R t}$ with corrections that decay
exponentially with $d$, the radius of the region $R$. In the figure,
the region $M$ represents a membrane of constant thickness surrounding
the region $R$.
\label{LR}}
\end{figure}

\begin{figure}
\includegraphics[width=.5\textwidth]{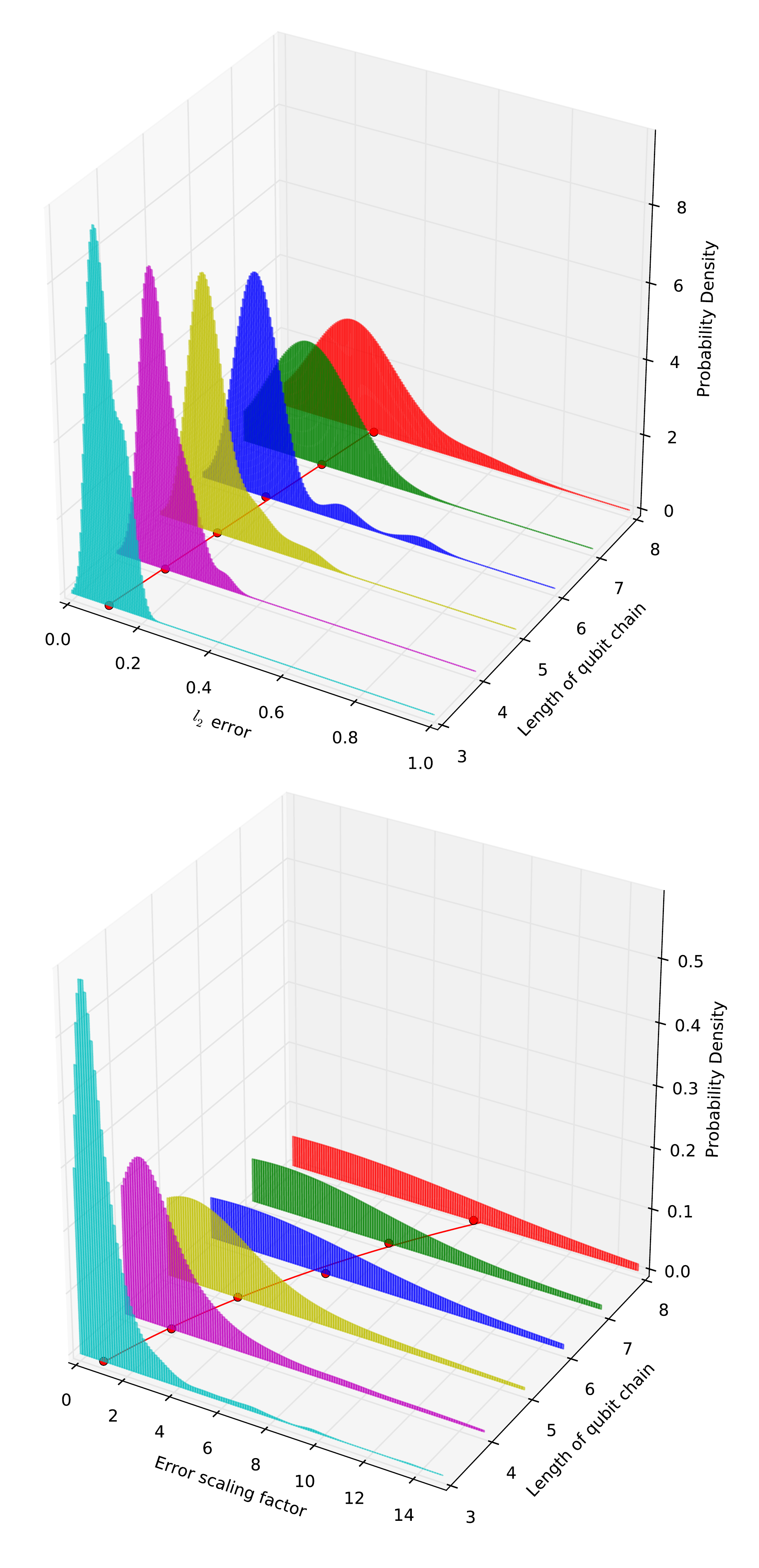}
\caption{
Error in estimates of the parameters of local Hamiltonians.
The systems consist of linear chains of qubits with randomly chosen
2-local Hamiltonians $\op H$---each coefficient has norm uniformly
distributed between 0.8 and 1.2.  Starting in an initial product
state, the system is evolved for $t=10^{-3}$, and the expectation of
randomly chosen observables is measured with precision $\epsilon$. The
resulting linear constraints Eq.~\eqref{eq:linear_dynamics} are solved
using Moore-Penrose pseudo-inverse to obtain an estimated Hamiltonian
$\tilde H$.  (Top) Distribution of the error $\frac 1d\sqrt{\tr(\op H
  - \tilde H)^2}$ over different realization of the random Hamiltonian
for $\epsilon = 10^{-4}$.  The red dots correspond to the mean
distance and the solid lines is a linear fit.  (Bottom) Distribution
of error scaling factors---i.e. the factor by which the measurement
accuracy $\epsilon$ is amplified when computing the pseudo-inverse.
The red dots indicate the average error scaling factor for each chain
length (the red line is a quadratic fit).
    \label{fig:rms}}
\end{figure}

The characterization of local Hamiltonians and Lindbladians relies
heavily on the Lieb-Robinson bound~\cite{LR72a,Has04} that shows that a
local Hamiltonian generates a causal evolution, with effects
propagating at a finite velocity $v$ (note that this bound has been
generalized to the setting of dissipative systems~\cite{P10a}, so our
derivation holds for local Lindbladians as well). A local
Hamiltonians acting on $n$ particles is of the
form $\hat H = \sum_X \hat H_X$ where $X$ labels subsets of $n$
particles, each term has bounded norm $\|\hat H_X\| \leq E$, and acts
on at most $k$ neighboring particles, such that $H_X = 0$ when $|X| >
k$. The evolution of an operator is governed by the equation
$\frac\partial{\partial t} \hat A(t) = i[\hat H, \hat A]$.  Break the
Hamiltonian into $\hat H = \hat H_0 + \hat H_M$, where $\hat H_M$
contains all the terms $\hat H_X$ that intersect a membrane $M$
surrounding the operator $\hat A$ (see Fig.~\ref{LR}).  The
idea of
this membrane is to disconnect its interior, denoted region $R$, from
the rest of the particles. Indeed, $e^{i\hat H_0 t}\hat A e^{-i\hat
  H_0t} = e^{i\hat H_R t}\hat A e^{-i\hat H_Rt}$ where $\hat H_R$ in
the Hamiltonian acting only inside the membrane (see
Fig.~\ref{LR}). The differential equation for $\hat A(t)$ is
\begin{equation}
\frac{\partial}{\partial t} \hat A(t) = i[\hat H_0, \hat A(t)] +i[\hat H_M,\hat
A],
\end{equation}
which has solution
\begin{align}
\hat A(t) =& e^{i\hat H_0 t} \hat A(0) e^{-i\hat H_0 t} \nonumber \\
&+ i\int_0^t e^{i\hat H_M (t-s)}[\hat H_M, \hat A(s)] e^{-i\hat H_M (t-s)} ds \\
=& e^{i\hat H_R t} \hat A(0) e^{-i\hat H_R t} \nonumber \\
&+ i\int_0^t e^{i\hat H_M (t-s)}[\hat H_M, \hat A(s)] e^{-i\hat H_M (t-s)} ds
\end{align}
as can be verified directly by differentiation. The commutator appearing in the
second term can be bounded by
\begin{equation}
\| [\hat H_M, \hat A(s)]\| 
\leq c V \|\hat A\| \| \hat H_M \| \exp\left(-\frac{d-vt}{\xi}\right)
\end{equation}
where $V$ is the number of sites in the support of the observable
$\hat A$, and $c$, $v$, and $\xi$ are constant that depend only on the
microscopic details of the system, independent of the system
size. This is known as the the Lieb-Robinson bound.  Integrating, we
obtain
\begin{align}
&\|\hat A(t) - e^{i\hat H_R t}\hat A(0) e^{-i\hat H_R t}\| \\
& \leq c t V \|\hat A\| \|\hat H_M\| \| \exp\left(-\frac{d-vt}{\xi}\right).
\nonumber
\end{align}
Expanding the exponential to first order yields
\begin{align}
&\|\hat A(t)  - \hat A(0) -it[\hat H_R ,A(0)] \| \\
& \leq c t V \|\hat A\| \|\hat H_M\| \| \exp\left(-\frac{d-vt}{\xi}\right)
\nonumber\\
& + c' \|\hat A\| \|\hat H_R\|^2 t^2. \nonumber
\end{align}
Because $\hat H_R$ and $\hat H_M$ represent respectively the
Hamiltonian of a ball of radius $d$ and the Hamiltonian for a constant
thickness membrane around that ball, they grow proportionally to $d^D$
and $d^{D-1}$ respectively, where $D$ is the spatial dimension,
\emph{i.e.}, $\|\hat H_R\| \leq \alpha d^D$ and $\|\hat H_M\| \leq
\alpha d^{D-1}$ for some constant $\alpha$.  Choosing $d \approx vt +
\log(cV/c't)$ such that
\begin{equation}
d^{D+1} \exp\left(\frac d\xi \right) \geq \frac{cV}{c't}
\exp\left(\frac{vt}\xi\right) ,
\end{equation}
we obtain 
\begin{align}
&\|\hat A(t) - \hat A(0)-  it[\hat H_R ,\hat A(0)] \|  \leq \kappa \|\hat A\|
\left[vt +\log\left(\frac{cV}{c't}\right)\right]^2 t^2
\end{align}
for some constant $\kappa = 2c'\alpha^2$. 

For a short time $t$, the expectation value of any observable
$\hat A$ evolves as
\begin{equation}
\langle \hat A(t)\rangle_{\hat
  \rho} - \tr \hat A\hat \rho = it\langle[\hat H,\hat A]\rangle_{\hat
  \rho} + \cO(\|\hat H\|^2 t^2).
  \label{eq:linear_dynamics}
\end{equation}
 By experimentally measuring this
expectation value, we obtain one linear constraint on the Hamiltonian.
Varying over different observables $\hat A_i$ and initial states $\hat
\rho_j$, we obtain more linear constraints that we can write as
$W_{ij} = \langle \hat A_i(t)\rangle_{\hat \rho_j} - \tr \hat A_i\hat
\rho_j = it\langle[\hat H,\hat A_i]\rangle_{\hat \rho_j}$ where we
have dropped the higher order terms $\cO(\|\hat H\|^2 t^2)$. Writing
$\hat H$ in an operator basis $\hat H = \sum_l h_l \hat P_l$, we
obtain the linear equation 
\begin{equation}
W_{ij} = \sum_l T_{ij,l} h_l
\label{eq:LI}
\end{equation}
 where
$T_{ij,l} = it \tr \hat \rho_j [\hat P_l,\hat A_i]$. The Hamiltonian
can be learned by inverting this linear equation~\cite{SML+10}.
  
There are in general four important caveats to this approach: {\em 1)}
the evolution time $t$ must be extremely short $t\ll \|H\|^{-1}$,
going to 0 as the number of particles grows; {\em 2)} there are
exponentially many $h_i$ to learn; {\em 3)} there are exponentially
many observables $\hat A_k$ and initial states $\hat\rho_j$ to be
measured and prepared experimentally; and {\em 4)} the quantities $\tr
\hat A\hat \rho$ and $\langle[\hat H,\hat A]\rangle_{\hat \rho}$ can
be exponentially difficult to compute. Based on
Eq.~\eqref{eq:linear_dynamics}, all these problems disappear when the
Hamiltonian is {\em local} as described in the main text.

Numerical experiments were performed for local Hamiltonians, and the
results are plotted in Fig.~\ref{fig:rms}. The systems we considered
were small chains of qubits with random nearest neighbour
interactions. The system evolution was calculated exactly for a short
amount of time, and the linearized problem was inverted using the
Moore-Penrose pseudoinverse. Since these Hamiltonians are drawn at
random (but with maximum strength for each term independent of the
system size), we calculate the average $l_2$ distance between the
estimated Hamiltonian and the actual Hamiltonian (top of
Fig.~\ref{fig:rms}), as well as the quantiles for error propagation
scaling factor of each of the elements of $h_l$, given by $\sum_{ij}
|T_{ij,l}^+|^2$ (bottom of Fig.~\ref{fig:rms}).  The results clearly
indicate well behaved error scaling for these systems, even under
finite statistical error in the estimation of observable expectations.

\end{document}